\documentclass[a4paper,10pt,twocolumn,aps]{revtex4-1}

\usepackage{amsmath}
\usepackage{amssymb}
\usepackage{graphicx}
\usepackage{epstopdf}
\usepackage{natbib}

\usepackage{bbm}
\usepackage{bm}
\usepackage{dcolumn}
\providecommand{\ket}[1]{|#1\rangle}
\providecommand{\bra}[1]{\langle#1|}

\providecommand{\idop}{\mathbbm 1}

\begin{document}
\title{Two-qubit gates for decoherence-free qubits using a ring exchange interaction}%
\author{Bobby Antonio}
\email[email: ]{bobbyantonio@gmail.com}
\author{Sougato Bose}
\affiliation{University College London, Gower Street, London, WC1E 6BT}
\date{\today}

\begin{abstract}
It is known that it is possible to encode a logical qubit over many physical qubits such that it is immune to the effects of collective decoherence, and it is possible to perform universal quantum computation using these `decoherence-free' qubits. However, current proposed methods of performing gates on these encoded qubits could be difficult to implement, or could take too much time to perform. Here we investigate whether exploiting ring-exchange interactions, which may be naturally present, can simplify the implementation of these gates in any way. Using a ring exchange interaction, we have found a way to create a controlled-Z gate on the 4-qubit decoherence-free subspace and the 3-qubit decoherence-free subsystem using a sequence with 5 pulses. This could be useful in situations where simplicity is important or where ring exchange interactions are prominent. We also investigate how timing errors and magnetic field fluctuations affect the fidelity of this gate.
\end{abstract}
\maketitle


\section{Introduction}\label{sec:Intro}

Quantum computers appear to offer a speed up compared to conventional classical computers, in problems such as factoring numbers~\cite{Shor1995} or searching unstructured databases~\cite{Grover1996}. One of the biggest obstacles to realising a useful, scalable quantum computer is making it robust against interactions with the environment that cause decoherence, in which information is irreversibly transferred from the system to the environment. There are two approaches to this problem: there is the `software' approach, where the logical qubit is encoded over several `physical' qubits in such a way that the errors due to the environment can be spotted and corrected easily; examples of this include the nine-qubit Shor code~\cite{Shor1997} and the 7-qubit Steane code~\cite{Steane1996}. The other approach is the `hardware' approach, in which the logical qubit is encoded over several physical qubits in such a way that the decoherence has minimal effect in the first place. One particular example of this is where we encode the logical information in part of the full Hilbert space in which the noise has no effect; such an encoding is called a `decoherence-free subspace' (DF subspace) or more generally a `decoherence-free subsystem' (DF subsystem)~\cite{ChuangYamamoto1995,DuanGuo1997,Lidar1998,ZanardiRasetti1997}.

It was shown in~\cite{Lidar1998,ZanardiRasetti1997,Bacon2000} that universal quantum computation can be performed inside a DF subspace or subsystem. To do this, we must be able to perform certain single qubit rotations as well as gates between two encoded qubits (such as a controlled-Z gate)~\cite{Barenco95}. For environmental noise which acts uniformly over all physical qubits (`collective decoherence'), explicit gate sequences which would realise this have been found in~\cite{Bacon2000,DiVincenzo2000,BaconThesis,Hsieh2003,Fong2011} for qubits encoded over 3 or 4 physical qubits (forming a 3-qubit DF subsystem and 4-qubit DF subspace, respectively). Performing single qubit rotations in the 4-qubit DF subspace and 3-qubit DF subsystem is relatively straightforward (see e.g.~\cite{BaconThesis,Hsieh2003}), however creating two-qubit gates is not; the methods found so far to perform two-qubit gates involve either interactions that could be difficult to create in an experiment~\cite{BaconThesis,Bacon2000}, or involve large numbers of gates to be switched on and off sequentially (e.g.\ 22 gates in 13 time steps for the 3-qubit DF subsystem in~\cite{Fong2011}), or use perturbative/complicated control sequences to create these gates~\cite{Jiang2009}.

Our aim is to simplify the existing methods of universal quantum computation in the 3-qubit DF subsystem and 4-qubit DF subspace, by finding alternative ways to perform 2-qubit gates using less operations, and using operations which are easier to implement (i.e.\ require less control), so that realisation of decoherence-free qubits might be more attainable in an experiment. We are also interested in seeing whether or not the presence of ring exchange interactions can lead to simpler gates on these encoded qubits. This work is partly motivated by the recent experimental advances in realising quadruple quantum dots in a square configuration~\cite{Thalineau2012}.

The paper is laid out as follows: In sec.~\ref{sec:Background} we introduce some important background material. In sec.~\ref{sec:Two} we discuss our main result constructing two qubit gates in the 3-qubit DF subsystem and 4-qubit DF subspace, and then in sec.~\ref{sec:Performance} we test the performance of these gates when coupling errors or magnetic fluctuations occur.

\section{Background information}\label{sec:Background}

\subsection{Decoherence-free subspaces and subsystems}

Consider a number of qubits coupled to an environment; when talking about these qubits we make an important distinction between \emph{encoded} qubits and \emph{physical} qubits; an encoded qubit is defined over several physical qubits (e.g.\ we will be using encodings later on which yield one encoded qubit for 3 or 4 physical qubits). The Hamiltonian for this system of qubits together with an environment can be written
\begin{align}
H = H_{sys} \otimes \idop_E + \idop_{sys}\otimes H_{E} + H_I,
\end{align}
where $H_{sys}$, $H_E$ and $H_I$ are the system, environment and system-environment interaction Hamiltonians, respectively. We can generically write the interaction Hamiltonian as
\begin{align}
H_I = \sum_{\alpha} A_{\alpha} \otimes B_{\alpha},
\end{align}
where $A_{\alpha}$ and $B_{\alpha}$ are some operators acting on the system and environment, respectively. For certain choices of $A_{\alpha}$ and $B_{\alpha}$, certain sets of states are transformed identically by these interactions, and so by encoding our information in these states we can reduce the effect of the environment to multiplication by a global phase, which goes unnoticed; such an encoding is called a \emph{decoherence-free subspace}. More generally, we can encode the information in a set of states such that the environment acts non-trivially on these states, but such that these transformations only couple to certain degrees of freedom which do not change the stored information; such an encoding is called a \emph{decoherence-free subsystem}. The particular encoding required will depend on the interactions, and in this paper we consider a particular kind of system-environment interaction for which it is possible to construct a DF subspace or subsystem: \emph{collective decoherence}. Collective decoherence occurs when each physical qubit interacts identically with the environment (e.g.\ by being close enough together with respect to the environment). In the case of collective decoherence acting on $N$ qubits, we can then write $H_I$ as
\begin{align}
H_I = \sum_{\alpha = x,y,z} S_{\alpha} \otimes B_{\alpha},
\end{align}
where $S_{\alpha} := \sum_{n=1}^N \sigma^{\alpha}_{n}$, and $\sigma^{\alpha}_{n}$ is a Pauli operator acting on the $n^{th}$ physical qubit, with $\alpha = x,y,z$. This is referred as \emph{strong} collective decoherence in the literature, and we will assume that there is strong collective decoherence acting on our encoded qubits for the rest of this paper, unless stated otherwise. For a more in depth discussion of DF subspaces and subsystems, the reader is referred to~\cite{Lidar1998,Lidar2003,BaconThesis,DuanGuo1998,PalmaSuominen1996,Kempe2001}.

The 3-qubit encoding which we use in this paper, and which has been used previously in~\cite{DiVincenzo2000,Fong2011}, acts as a DF subsystem for strong collective decoherence; it is defined over four eigenstates of the operators $S^2 = (S_x)^2 + (S_y)^2 + (S_z)^2$, $S_z$ and $S^2_{1,2}$, where $S^2_{1,2}$ is the total spin operator acting only on physical qubits $1$ and $2$. These three operators have eigenvalues $m_S(m_S + 1)\hbar^2$, $m_z$ and $m_{1,2}(m_{1,2} + 1)\hbar^2$ respectively, and we label the states according to these three quantum numbers:
\begin{align}
&\ket{\bar{0}^{(3)}_{+1}}=\ket{m_S = 1/2, m_z = 1/2, m_{1,2} = 0 } = \ket{\psi^-}_{12}\ket{0}_3\nonumber\\
&\ket{\bar{0}^{(3)}_{-1}}=\ket{1/2,-1/2,0} = \ket{\psi^-}_{12}\ket{1}_3\nonumber\\
&\ket{\bar{1}^{(3)}_{+1}}=\ket{1/2,1/2,1} = \frac{1}{\sqrt{3}}( \sqrt{2} \ket{T_+}_{12}\ket{1}_3 - \ket{T_0}_{12}\ket{0}_3)\nonumber\\
&\ket{\bar{1}^{(3)}_{-1}}=\ket{1/2,-1/2,1} = \frac{1}{\sqrt{3}}( \ket{T_0}_{12}\ket{1}_3 - \sqrt{2} \ket{T_{-}}_{12}\ket{0}_3).
\end{align}
Here we have used the singlet states on qubits $i$ and $j$, defined as $\ket{\psi^-}_{ij} := (\ket{01}_{ij} - \ket{10}_{ij})/\sqrt{2}$ and the triplet states $\ket{T_+}_{ij} = \ket{00}_{ij}$, $\ket{T_-}_{ij} = \ket{11}_{ij} $,  $\ket{T_0}_{ij} = (\ket{01}_{ij} + \ket{10}_{ij})/\sqrt{2}$. We define the logical zero state $(\ket{\bar{0}^{(3)}})$ in this 3-qubit subsystem to be an arbitrary combination of the first two states, i.e. $\ket{\bar{0}^{(3)}} := \zeta \ket{\bar{0}^{(3)}_{+1}}   + \gamma \ket{\bar{0}^{(3)}_{-1}} $, whilst the logical one state $(\ket{\bar{1}^{(3)}})$ is a superposition of the last two states with the same coefficients, $\ket{\bar{1}^{(3)}} := \zeta \ket{\bar{1}^{(3)}_{+1}} + \gamma \ket{\bar{1}^{(3)}_{-1}}$. The action of collective decoherence on this encoding can change the values of $\zeta$ and $\gamma$, but it will not couple states with different values of ${m}_{1,2}$, so the information is preserved. The arbitrary choice of $\zeta$ and $\gamma$ is called a gauge degree of freedom, and any transformation which only changes the values of $\zeta$ and $\gamma$ is called a gauge transformation (in this case a gauge transformation is an operation which only changes the value of $m_Z$). In addition we refer to the separate subspaces with different $m_Z$ values as {\it gauge subspaces}.

The 4-qubit encoding we use in this paper acts as a DF subspace for strong collective decoherence, and is in fact the smallest number of qubits over which it is possible to encode a strong DF subspace~\cite{Lidar2003}. The logical subspace $\{ \ket{\bar{0}^{(4)}},\ket{\bar{1}^{(4)}} \}$ 
is defined over the two states $\ket{ m_S = 0, m_z=0, m_{1,2} = m_{3,4} = 0}$ and $\ket{ m_s= 0, m_z =0,m_{1,2} = m_{3,4} =1}$:
\begin{align}\label{eqn:States}
\ket{\bar{0}^{(4)}} & :=  \ket{\psi^-}_{12} \ket{\psi^-}_{34}\nonumber\\
\ket{\bar{1}^{(4)}} & :=  \frac{1}{\sqrt{3}} \left[ \ket{T_+}_{12} \ket{T_-}_{34} - \ket{T_0}_{12} \ket{T_0}_{34}\right.\nonumber\\ 
&\left.+ \ket{T_-}_{12} \ket{T_+}_{34}\right].
\end{align}
The 4-qubit encoding has one additional desirable property; it also functions as a \emph{supercoherent qubit}~\cite{Bacon2001}. Supercoherence would allow resistance to errors acting on individual physical qubits, with a mechanism as follows: When an error along any direction is applied to the physical qubits in the $m_S=0$ states in eqn. (\ref{eqn:States}), it is accompanied by a change in the $m_S$ value by 1~\cite{Bacon2001}. In order to use this to create a supercoherent qubit, we could switch on the Hamiltonian ${H}_{SC}$, defined as
\begin{equation}\label{eqn:Hsc}
{H}_{SC} = J_{SC}\sum_{i,j} {E}_{ij},
\end{equation}
where ${E}_{ij} :=  {\sigma}^i_{x}{\sigma}^j_{x} +{\sigma}^i_{y}{\sigma}^j_{y} + {\sigma}_i^{z}{\sigma}_j^{z}$,  and the sum is over all pairs of the 4 physical qubits. With this Hamiltonian switched on, the $m_S=0$ states are degenerate and lowest in energy, with an energy gap between the $m_S=0$ states and any other states. Thus any decoherence process acting on individual physical qubits in the $m_S = 0$ state involves an increase in energy of the encoded qubit, and will lead to a transfer of energy from the environment to the system, which we can inhibit by cooling the environment. Thus supercoherent qubits would be very useful as quantum memories, and it was argued in~\cite{Bacon2001} that computation with supercoherent qubits could be performed provided the interaction strength between qubits was small enough compared to $J_{SC}$ (leading to a trade-off between the speed of operations and the robustness against errors). In this paper, we will not aim to make our interactions supercoherent as well (i.e.\ we envisage a protocol in which we use the supercoherent mechanism as a means to reliably store information, but turn off the supercoherent Hamiltonian ${H}_{SC}$ when we interact encoded qubits). 

\begin{figure}[h]
\includegraphics[scale=0.4]{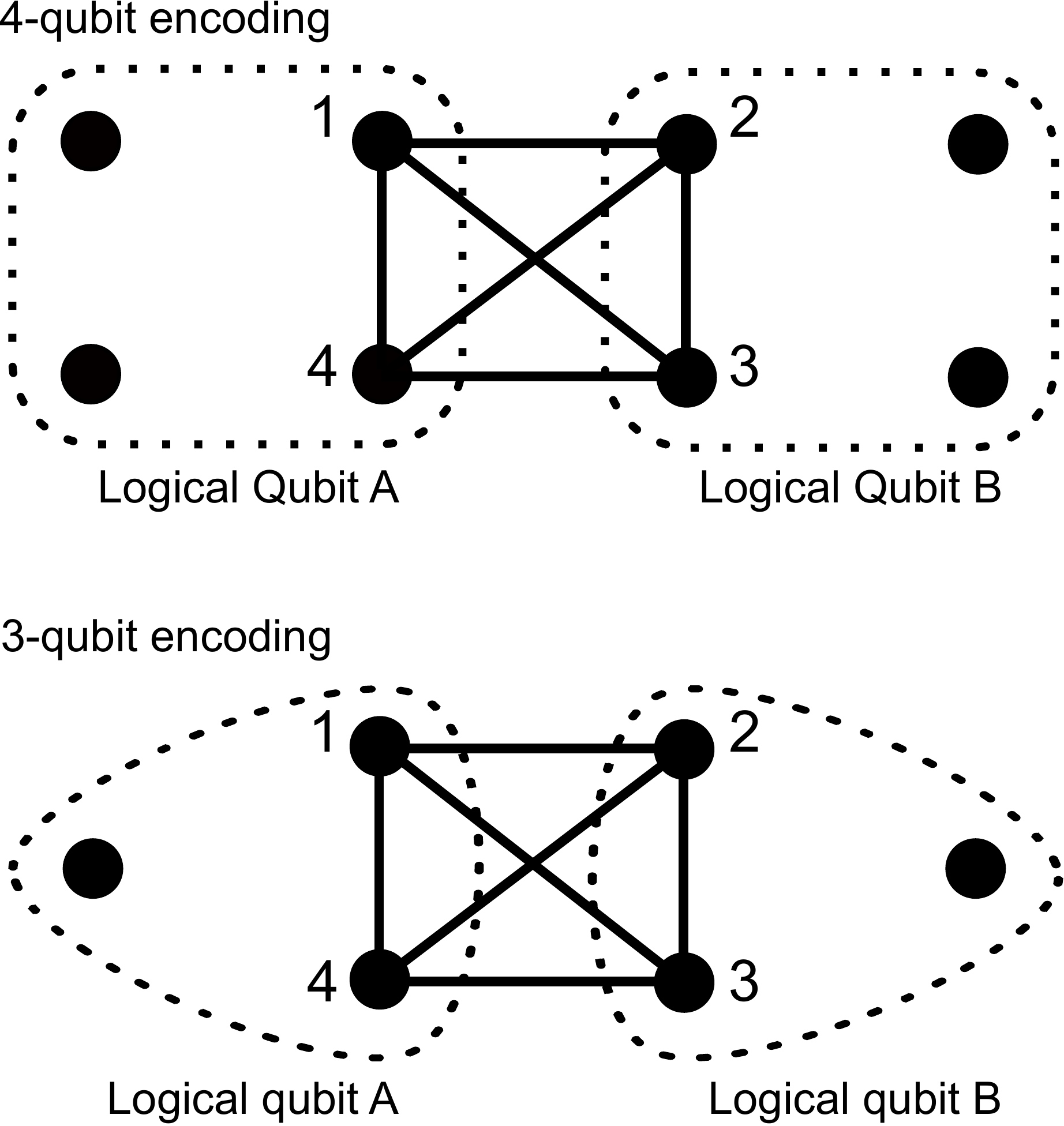}
\caption{An illustration of the two geometries of qubits we consider when constructing our logical qubits. Filled circles represent physical qubits, and solid lines illustrate the kind of interactions we consider when looking for a gate in this paper. The top diagram shows the layout for the 4-qubit encoding and the bottom diagram shows the layout for the 3-qubit encoding. }
\label{fig:Plaquette}
\end{figure}

For this investigation, we consider physical qubits arranged in a regular formation; three spins in an equilateral triangle for the 3-qubit DF subsystem, and four spins in a square for the 4-qubit DF subspace. The interactions we consider for both encodings are interactions between the middle 4 spins (see Fig.~\ref{fig:Plaquette} for an illustration of this). 

\begin{figure}[h]
\includegraphics[scale=0.4]{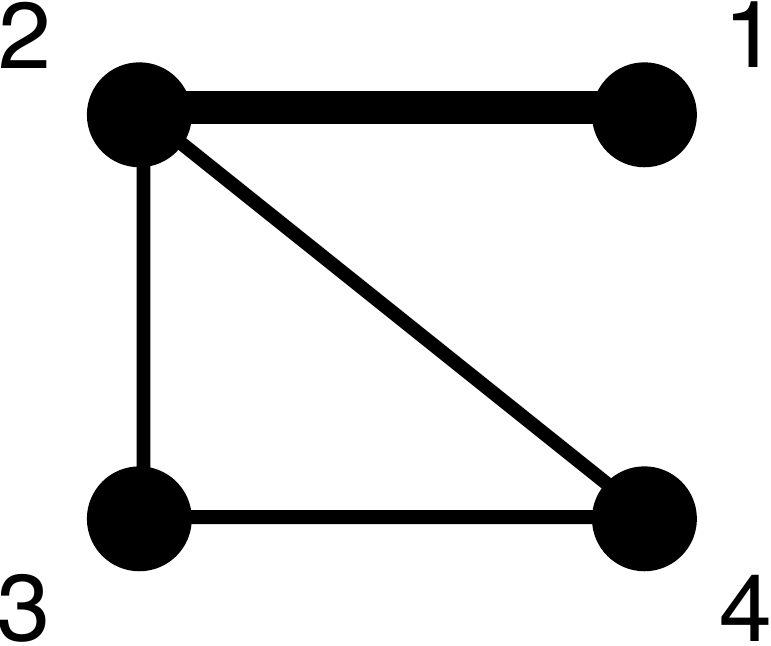}
\caption{An illustration of interactions used in~\cite{BaconThesis} to interact two 4-qubit encoded qubits, which could be challenging to implement in an experiment. Filled circles represent physical qubits, and solid lines represent exchange interactions. Note that in~\cite{BaconThesis} these interactions are used when the qubits are taken out of the logical subspace into the larger 14-dimensional singlet subspace over 8 qubits, and so this is not simply a local unitary transformation.}
\label{fig:Bacon}
\end{figure}

As an example of why we might want to simplify the existing two-qubit gates in the 4-qubit DF subspace, see Fig.~\ref{fig:Bacon} for an illustration of one such interaction used in~\cite{BaconThesis,Bacon2000} to create a gate. The Hamiltonian for this interaction is:
\begin{align}
H = 3 {E}_{12} + \frac{2}{3}({E}_{24} + {E}_{23} + {E}_{34}).
\end{align}
Given the large difference in couplings between qubits 1 and 2 compared to the other qubits, and the absence of coupling between qubits 1 and 4, this could be a challenging gate to realise (note that although this only acts within one encoded qubit, this Hamiltonian is turned on whilst we are out of the logical subspace, and so is not simply a local unitary operation).


\subsection{Ring Exchange Interactions} \label{sec:Rexchange}
When constructing our gate, we will include `ring exchange' interactions. These interactions, which are used to explain  excitations in $La_2 Cu O_4$~\cite{Katanin02,Coldea01} and become important in electrons forming a Wigner crystal~\cite{Bernu2001,Voelker2001}, appear as corrections in the exchange Hamiltonian due to higher order hopping processes between different physical qubits in the extended Hubbard Hamiltonian~\cite{Takahashi1977}. They have also been investigated in the context of quantum computing~\cite{Scarola2005,Mizel2004,Mizel2004B}, and it is clear from these papers that ring exchange processes should not be ignored.  Including ring exchange terms, the modified Hamiltonian for 4 spin-1/2 particles located at sites 1,2,3 and 4 becomes~\cite{Takahashi1977}:
\begin{align}\label{Rexchange1}
{H}&= \sum_{i\ne j}  J_{ij} {E}_{ij} \nonumber \\
 &+C_{1234} [ {E}_{12} {E}_{34} + {E}_{14} {E}_{23}  -  {E}_{13} {E}_{24}] \nonumber \\
  &+ C_{1324} [ {E}_{13} {E}_{24} + {E}_{14} {E}_{23}  -  {E}_{12} {E}_{34}]\nonumber\\
 &+C_{1342} [ {E}_{13} {E}_{24} + {E}_{12} {E}_{34}  -  {E}_{14} {E}_{23}] \nonumber\\
 & + \mathcal{O} \left(\frac{t^5}{U^4} \right),
\end{align}
where $t$ represent the average tunnelling coefficient between electron sites, $U$ is the on-site Coulomb interaction, $J_{ij}$ is the exchange coupling between sites $i$ and $j$, $E_{ij}$ is the exchange interaction defined in sec.~\ref{sec:Background}, and $C_{ijkl}$ are ring exchange coefficients. The ring exchange coefficients and exchange coupling are linked, since they both rely on $t_{ij}$, the tunnelling coefficient between electron sites $i$ and $j$. More specifically, $J_{ij} \propto t_{ij}$ and $C_{ijkl} \propto t_{ij} t_{jk} t_{kl} t_{li}$ where none of the indices $i,j,k,l$ are equal. This means that these ring exchange terms will only appear when there are exchange terms present which form a loop (i.e.\ when a physical qubit is indirectly coupled to itself). For example, if we have four electrons with exchange interactions between electrons 1-2, 2-3, 3-4, and 4-1, then there will be ring exchange terms present. In~\cite{Scarola2005}, they also find that the presence of magnetic fields changes the coefficients $C_{ijkl}$ and introduces three-body terms, with couplings that depend on the magnetic flux passing through 3 or 4-site loops. Here we assume that these magnetic fields are low enough so that the magnetic flux has negligible impact and three body terms can be ignored, and we leave the effects of larger magnetic fields to later work.

With four qubits arranged on a square, it is possible to make all of the exchange couplings uniform, and since this will simplify things considerably, whenever we need to include ring-exchange terms in this paper we will assume that this is the case. Taking uniform interactions in eqn. (\ref{Rexchange1}) results in a `symmetric' version of the Hamiltonian, ${H}_S$, which takes the form derived in~\cite{Mizel2004}:
\begin{align}\label{Rexchange2}
{H}_S &= J_{\Box}{H}_{\Box} + J_{\times} {H}_{\times} + J_{\circlearrowleft} {H}_{\circlearrowleft} \nonumber\\
&= J ({H}_{\Box} + {H}_{\times}) + J_{\circlearrowleft} {H}_{\circlearrowleft} \nonumber\\
 &= J \sum_{n=1}^4 {E}_{n,n+1} +J \sum_{n=1}^2 {E}_{n,n+2} \nonumber \\
 &+ J_{{\circlearrowleft}} [{E}_{12} {E}_{34} + {E}_{14} {E}_{23}  + {E}_{13} {E}_{24}]\nonumber\\
 &=J \left( {H}_{\Box} + {H}_{\times} + \alpha {H}_{\circlearrowleft} \right),
\end{align}
where ${H}_{\Box}$, ${H}_{\times}$, ${H}_{\circlearrowleft}$ represent the nearest-neighbour, next-nearest-neighbour and ring-exchange Hamiltonians, respectively, and $J_{\Box}$, $J_{\times}$, $J_{\circlearrowleft}$ are the corresponding interaction strengths for these Hamiltonians. Since we have equal coupling between all sites, $J_{\Box} = J_{\times} = J$ in this equation, and in the final line we have defined $\alpha := J_{\circlearrowleft} /J_{\Box} \equiv  J_{\circlearrowleft} /J$ as the ratio of ring exchange terms to the nearest/next-nearest neighbour terms. Restricting ourselves to symmetric Hamiltonians of this form simplifies things considerably; ${H}_{\circlearrowleft}$ commutes with ${H}_{\Box}$ and ${H}_{\times}$, and as argued in~\cite{Mizel2004}, this form of Hamiltonian contains enough degrees of freedom to fix all of the eigenvalues, and so we do not need to take into account any higher order terms, unlike the perturbative expansion in eq. (\ref{Rexchange1}). Since single qubit rotations of the encoded qubit can be performed by exchange interactions between physical qubits, creating a two-qubit interaction comes down to simulating a 4-body interaction of the form $E_{ij}E_{kl}$, and since the ring exchange interactions contain terms similar in nature to these, it seems plausible that we can use these to simplify the 2-qubit gates acting on the encoded qubits.

\section{Two-qubit gates}\label{sec:Two}

We now look at creating a controlled-${Z}$ gate ($CZ$) between two encoded qubits, which, along with certain single qubit gates (e.g. Hadamard and $R_z(\pi/4)$ gates), enables us to perform universal quantum computation \cite{Barenco95}. An important point to note is that, in this case if we can find a two-qubit gate which works for the 3-qubit DF subsystem, this is also a valid gate for the 4-qubit DF subspace (see appendix~\ref{App1}), so we only need to search for one pulse sequence for both of these encodings. 
Performing two qubit gates is, predictably, not as straightforward as single gates, mainly because we move into a much larger Hilbert space as soon as interactions between the two qubits are turned on. Since $[ {S}^2,{E}_{ij}] = 0$ (provided $i$,$j$ are both qubits that ${S}^2$ operates on), if we start with two encoded qubits in the logical subspace with a total spin number $S$, then when we interact them together using exchange interactions they will still have overall spin number $S$, so we can use this property to confine ourselves into a region of the full Hilbert space, to speed up calculations.

We also constrain ourselves to couplings which are realistically possible; we only couple sites which can realistically be placed near each other (by e.g.\ placing two encoded qubits side-by-side, see Fig.~\ref{fig:Plaquette}). The method we used to search for a quantum gate, once we had chosen a certain set of interactions to, uses the invariant quantities found by Makhlin~\cite{Makhlin2002}. In this paper, two invariant quantities $m_1$ and $m_2$ are derived for 2-qubit operations. Given a $4\times 4$ matrix ${M}$, we first transform ${M}$ into the Bell basis, ${M}\rightarrow {M}_B = {Q}^{\dagger}{M} {Q}$ where
\begin{align}
{Q} = \frac{1}{\sqrt{2}} \left( \begin{array}{c c c c} 1 &0& 0& i \\ 0 &i &1 &0 \\ 0 &i &-1& 0\\ 1& 0& 0& -i \end{array} \right).
\end{align}
Then defining $m = {M}_B^{T}{M}_B$, the two invariant quantities $m_1$ and $m_2$ are given by:
\begin{align}
&m_1({M}) = (\textrm{tr}\; m)^2/16\det {M}^{\dagger}\\
&m_2({M}) =((\textrm{tr}\; m)^2 -\textrm{tr}(m^2) )/4\det {M}^{\dagger}.
\end{align}
We can find these two invariant values for any operator, and if they both match with the value for a $CZ$ gate then the two gates are equivalent apart from some local operations. 

\begin{figure}[b]
\includegraphics[scale=0.2]{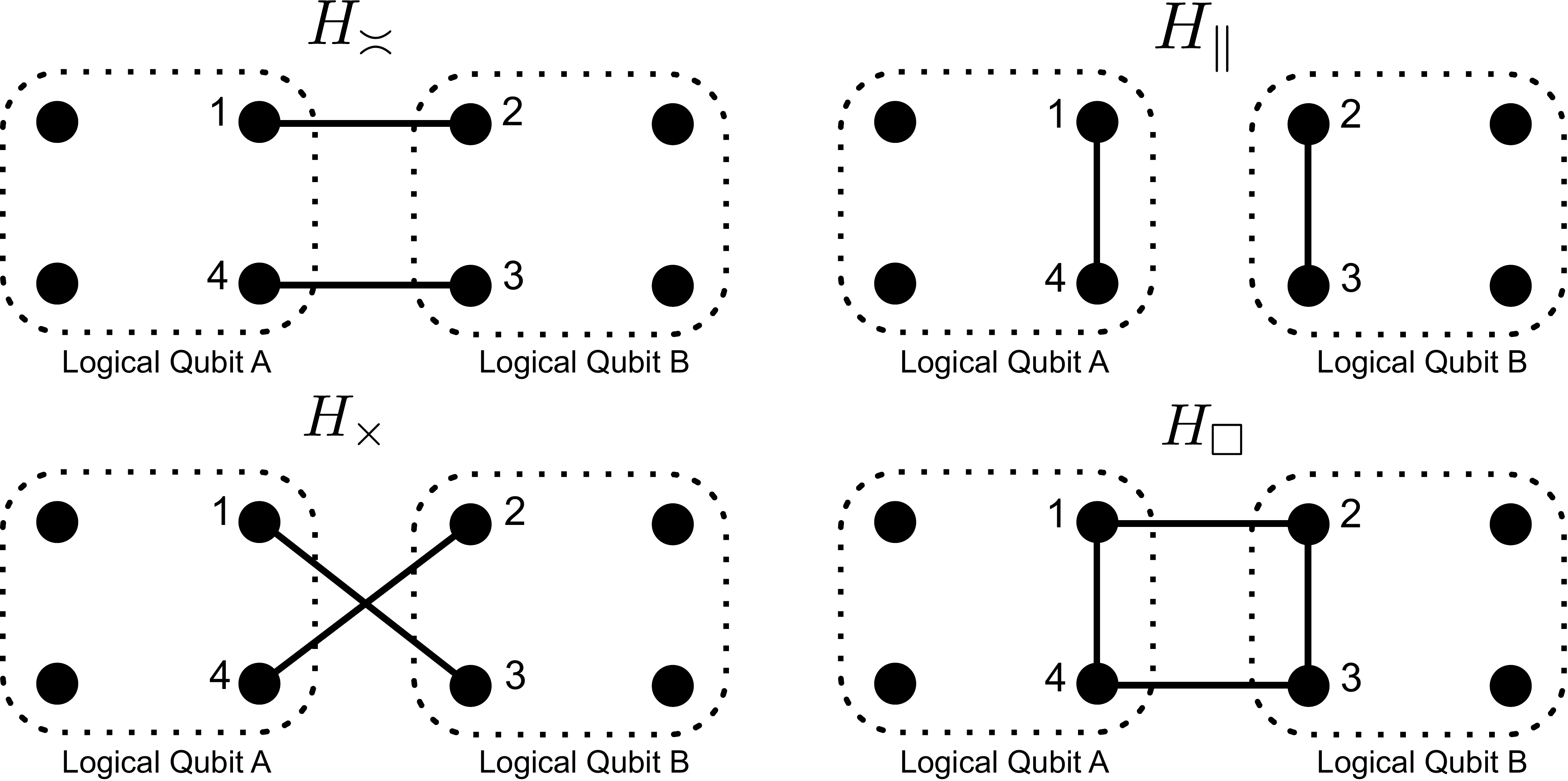}
\caption{An illustration of the interactions we use to construct the gates, excluding ring exchange interactions which are difficult to represent in this form. Filled circles represent physical qubits, and solid lines represent exchange interactions.}
\label{fig:AllHams}
\end{figure}

The Hamiltonians that we used to construct a gate with are the following:
\begin{align} \label{HamSeq}
{H}_{\asymp} &:= {E}_{12} + {E}_{34}, \hspace{2mm} {H}_{\parallel} := {E}_{14} + {E}_{23}\nonumber\\
{H}_{\times} &:= {E}_{13} + {E}_{24}, \hspace{2mm}{H}_{\Box} := {H}_{\asymp} + {H}_{\parallel}\nonumber\\
{H}_{\circlearrowleft} &:={E}_{12} {E}_{34} + {E}_{14} {E}_{23}  + {E}_{13} {E}_{24} 
\end{align}
The convention for numbering the qubits is shown in Fig.~\ref{fig:AllHams}, along with an illustration of these Hamiltonians (excluding $H_{\circlearrowleft}$ which is hard to represent in pictorial form). For each Hamiltonian in this set we define a corresponding unitary operator:
\begin{align}
 {U}_{n} &= \exp \left(-i \frac{J_n  {H}_{n} \tau_n}{ \hbar} \right) \equiv \exp \left(-i  {H}_{n} \theta_n \right)\\
 & \mbox{e.g.} \;  {U}_{\asymp} = \exp(-i  {H}_{\asymp} \theta_{\asymp}), \; \mbox{etc.} \nonumber
\end{align}
where $\theta_n := J_n \tau_n / \hbar$. Note that all of the $\theta_n$ are phases, so we are free to add multiples of $2 \pi$ without changing the properties of the gate. We then multiply all of these unitaries together in some order to get the full gate operator:
\begin{align}
{U}_{tot} = \prod_n {U}_n = \prod_n \exp(-i  {H}_{n} \theta_{n}).
\end{align}
Before comparing this to a $CZ$ gate, we must first project into the logical subspaces of the 3- or 4-qubit encodings, so $U_{tot} \to U'_{tot} = P_j^{\dagger} U_{tot}P_j$, where $P_j$ is a projector into the logical subspace $\mathcal{L}_j$ of the two encoded qubits, with the subscript $j$ indicating whether it is the 3-qubit encoding ($j=3$) or the 4-qubit encoding ($j=4$). We define these projection operators as
\begin{align}
&P_4 = \sum_{x,y \in \{0,1\}} \ket{\bar{x}^{(4)}}\ket{\bar{y}^{(4)}} \bra{\bar{x}^{(4)}}\bra{\bar{y}^{(4)}}\\
&P_3 = \sum_{i,j \in \{+1,-1\} } \sum_{x,y \in \{0,1\}} \ket{\bar{x}^{(3)}_{i}}\ket{\bar{y}^{(3)}_{j}}\bra{\bar{x}^{(3)}_{i}}\bra{\bar{y}^{(3)}_{j}}
\end{align}
Note that when combining two 3-qubit states together, there are four possibilities; three $S=1$ subspaces and one $S=0$ subspaces. Each of these subspaces is 4-dimensional, due to the gauge choices, so $P_3$ projects into a 16-dimensional subspace overall. We can define the projectors onto each of these 4-dimensional subspaces as
\begin{align}
&P_3^{(1,1)} = \sum_{x,y \in \{0,1\} } \ket{\bar{x}^{(3)}_{+1}}\ket{\bar{y}^{(3)}_{+1}}\bra{\bar{x}^{(3)}_{+1}}\bra{\bar{y}^{(3)}_{+1}}\nonumber\\
&P_3^{(1,0)} = \frac{1}{2}\sum_{i,j \in \{+1,-1\}}\sum_{x,y \in \{0,1\} } \ket{\bar{x}^{(3)}_{i}}\ket{\bar{y}^{(3)}_{-i}}\bra{\bar{x}^{(3)}_{j}}\bra{\bar{y}^{(3)}_{-j}}\nonumber\\
&P_3^{(1,-1)} := \sum_{x,y \in \{0,1\}} \ket{\bar{x}^{(3)}_{-1}}\ket{\bar{y}^{(3)}_{-1}}\bra{\bar{x}^{(3)}_{-1}}\bra{\bar{y}^{(3)}_{-1}}\nonumber\\
&P_3^{(0 ,0)} = \frac{1}{2} \sum_{i,j \in \{+1,-1\}}\sum_{x,y \in \{0,1\} } (-1)^{i+j}\ket{\bar{x}^{(3)}_{i}}\ket{\bar{y}^{(3)}_{-i}}\bra{\bar{x}^{(3)}_{j}}\bra{\bar{y}^{(3)}_{-j}}
\end{align}
In order to find the Makhlin invariant we must project into a 4-dimensional subspace; for the 3 -qubit encoding, we could project onto each of the $S=1$ subspaces and the $S=0$ subspace individually, and verify that the gate works in each subspace. However, this is not necessary, since $[\sum_{i=1}^N \sigma_i^\pm,E_{nm}] = 0$ if  $n,m \in \{ 1,2,...,N\}$, which means that if we have a $CZ$ gate which works in one of the $S=1$ subspaces, then this gate also works in any of the other two $S=1$ subspaces. So if we can find a gate which is locally equivalent to a $CZ$ in any one of the $S=1$ subspaces and also the $S=0$ subspace, it will be locally equivalent to a $CZ$ gate when acting on the overall 3-qubit DF subsystem up to some gauge transformation.

As mentioned before and shown in appendix~\ref{App1}, any parameters which work for the 4-qubit encoding will also work for the 3-qubit encoding (since we use exchange interactions restricted to the middle four physical qubits), so rather than searching for two separate parameters for the 4- and 3-qubit encodings, we can just search for gates which work for the 4-qubit encoding, which simplifies this process.
So we take the projected unitary $ U'_{tot} = P_4^{\dagger} U_{tot}P_4$, find the corresponding Makhlin invariants $m_1({U}'_{tot}),m_2({U}'_{tot})$, and compare these to the Makhlin invariants $m_1(CZ),m_2(CZ)$ of an ideal $CZ$ gate using the following function
\begin{equation}
f_m = \sum_{i=1}^2 \left|m_i(CZ) - m_i({U}'_{tot}) \right|,
\end{equation}
which gives us a measure of how close we are to a $CZ$ gate, excluding local unitary rotations. Minimising over $f_m$ will give us possible ways to implement a gate (or will tell us if it isn't possible for the type of interactions we are considering). The local operations required to transform our result to a $CZ$ gate will be easy to find compared to the difficulty of minimising over $f_m$, and so in this paper we focus on finding sequences which minimise $f_m$. We also need to consider how far out of the logical subspace we are, so we define the leakage parameter for both encodings as:
\begin{align}\label{eqn:leakage}
L_3 &:= 1 - \frac{1}{16}\| P_3 {U}'_{tot} P_3 \|^2\\
L_4 &:= 1 - \frac{1}{4}\| P_4 {U}'_{tot} P_4 \|^2
\end{align}
Where $\mathcal{L}$ denotes the logical subspace, $\| A\|$ is the Frobenius norm of $A$, and we divide by $16$ and $4$ respectively due to the different sizes of the logical spaces. Using these measures, we implemented a genetic algorithm (see e.g.~\cite{GeneticBook}), followed by a Nelder-Mead simplex search~\cite{NelderMead} once we had narrowed our search down to a sufficient level. The same method has been used in~\cite{DiVincenzo2000} and~\cite{Hsieh2003}. 

Finally we note an identity which makes this search over parameters easier to make. As noted in section~\ref{sec:Rexchange}, we consider ${H}_{\circlearrowleft}$ such that it commutes with ${H}_{\times}$ and ${H}_{\Box}$, which allows us to rearrange  ${U}_{\times} {U}_{\Box} {U}_{\circlearrowleft}$ as:
\begin{align}\label{eqn:Iden1}
 {U}_{\times} {U}_{\Box} {U}_{\circlearrowleft} &= e^{-i {H}_{\times} \theta_{\times}} e^{-i {H}_{\Box} \theta_{\Box}} e^{-i {H}_{\circlearrowleft} \theta_{\circlearrowleft}} \nonumber \\ 
&= e^{-i ({H}_{\times} \theta_{\times} + {H}_{\Box} \theta_{\Box} + {H}_{\circlearrowleft} \theta_{\circlearrowleft})} \nonumber\\
&= e^{-i ({H}_{\times} (\theta_{\times} - \theta_{\Box}))} e^{(-i  {H}_{\times}\theta_{\Box}  + {H}_{\Box}\theta_{\Box} + {H}_{\circlearrowleft}\theta_{\circlearrowleft} )} \nonumber\\
&= e^{-i ({H}_{\times} (\theta_{\times} - \theta_{\Box}))} e^{-i J\tau_{\Box}( {H}_{\times}  + {H}_{\Box} + (J_{\circlearrowleft}/J) {H}_{\circlearrowleft} )} \nonumber\\
&= e^{-i ({H}_{\times} (\theta_{\times} - \theta_{\Box}))} e^{-i \theta_{\Box} ( {H}_{\times}  + {H}_{\Box} + \alpha {H}_{\circlearrowleft} )} \nonumber\\
&= {U}_{\times}' {U}_S.
\end{align}
where
\begin{align}
{U}_S  = e^{-i[{H}_{\Box} + {H}_{\times} + \alpha {H}_{\circlearrowleft} ]} \equiv e^{-i {H}_S \theta_S}\nonumber\\
{U}_{\times}'  = e^{-i ({H}_{\times} (\theta_{\times} - \theta_{\Box}))}  =e^{-i {H}_{\times} \theta_{\times}'} 
\end{align}
In the above we set $\tau_{\Box} = \tau_{\circlearrowleft}$ since $H_{\Box}$ and $H_{\circlearrowleft}$ operate at the same time, and recall that $\alpha = J_{\circlearrowleft}/J $ and $J_{\Box} = J_{\times} = J$ in order to use the symmetric form of the ring exchange interaction, $H_{\circlearrowleft}$. Then using this identity, if we have within our gate sequential applications of the Hamiltonians ${H}_{\times}$, ${H}_{\Box}$ and ${H}_{\circlearrowleft}$ with parameters $\theta_{\Box}$, $\theta_{\times}$, $\theta_{\circlearrowleft}$ respectively, we can express this as the Hamiltonian in eq. (\ref{Rexchange2}) plus an additional next-nearest-neighbour term ${U}_{\times}' $ with parameter $\theta_{\times}'$ out in front (if $\theta_{\times}'$ is negative we can just add $2 \pi$ since it is just a phase). Also note that, since we have set $t_{\Box} = t_{\circlearrowleft}$, then $\alpha = \theta_{\circlearrowleft} / \theta_{\Box}$. This identity means that for the purposes of the minimisation we can treat ${H}_{\times}$, ${H}_{\Box}$ and ${H}_{\circlearrowleft}$ as if they were separate interactions with independent parameters and then combine them together at the end using the identity in eq. (\ref{eqn:Iden1}), and when we do combine them the value of $\alpha$ is set by $\theta_{\circlearrowleft} / \theta_{\Box}$.


\subsection{Results}

Starting with the simplest Hamiltonian which might result with ring exchange interactions, we tried many combinations of the Hamiltonians given above, performing a genetic search for each one followed by a Nelder-Mead search. Using the following sequence of interactions did not yield any gates with $f_m < 0.001$
\begin{align}
{U}_{1} &=  {U}_{\times} {U}_{\Box} {U}_{\circlearrowleft} \nonumber\\
{U}_{2} &= {U}_{\asymp} {U}_{\times} {U}_{\Box} {U}_{\circlearrowleft} \nonumber\\
{U}_{3} &= {U}_{\asymp}^{(1)} {U}_{\times} {U}_{\Box} {U}_{\circlearrowleft} {U}_\asymp^{(2)}.
\end{align}
However, the following combination produced several parameters which did work:
\begin{align} \label{Usequence}
{U}_{gate} &= {U}_{\asymp}^{(1)} {U}_{\parallel} {U}_{\times} {U}_{\Box} {U}_{\circlearrowleft} {U}_\asymp^{(2)} \nonumber\\
&= {U}_{\asymp}^{(1)} {U}_{\parallel} {U}_{\times}' {U}_S  {U}_{\asymp}^{(2)},
\end{align}
with corresponding parameter set:
\begin{equation}
\{\theta \} = \{ \theta_{\asymp}^{(1)}, \theta_{\parallel}, \theta_{\times}, \theta_{\Box}, \theta_{\circlearrowleft}, \theta_\asymp^{(2)} \},
\end{equation}
where the superscripts (1) and (2) are used to differentiate between the interactions used at the beginning and at the end of the gate. This is a gate which requires only $5$ separate pulses to perform (since we can use identity (\ref{eqn:Iden1})).

Using this combination of interactions, with parameters $\{ \theta \}$ given in table~\ref{thetaVals}, we were able to find a gate with a value of $f_m = O(10^{-16})$ and leakage $L = O(10^{-16})$ (i.e. both around the machine precision of $O(10^{-16})$, suggesting the existence of an exact solution). If we assume that for each interaction the coupling strength is limited to some maximum possible value $J_{max}$, then we can find the total gate time $T$ in units of $\hbar /J_{max}$, as an indicator of how long this gate would take compared to other gates. Note that we do not simply add the parameters in table~\ref{thetaVals}, since we apply the identity in eq. (\ref{eqn:Iden1}) first, so in fact the true gate time $T$ is
\begin{align}
T =\left( \sum_n \theta_n \right) - \theta_{\times} + (\theta_{\times}-\theta_{\Box})\textrm{mod}\;{2\pi}.
\end{align}
This gives a gate time of $16.7\; \hbar /J_{max}$. The gate in~\cite{Fong2011} has a total time of 9.9 in these units, and although we found other parameter sets which yielded similar gate times to this, we have picked the parameters with the most realistic ring exchange couplings (see sec.~\ref{sec:Constraints}).

\begin{table}[htbp] 
\caption{Parameters which realise a $CZ$ gate, up to local rotations.}
\centering
\begin{tabular}{>{\centering}m{0.2\linewidth} | >{\centering}m{0.4\linewidth}}
\hline \vspace{0.1cm}
  $\theta_{\asymp}^{(1)}$ &   2.748893584737 \tabularnewline\vspace{0.1cm}
  $\theta_{\parallel}$ & 4.319689917260 \tabularnewline\vspace{0.1cm}
  $\theta_{\times}$ & 2.552544025744  \tabularnewline\vspace{0.1cm}
  $\theta_{\Box}$ & 3.730678055907  \tabularnewline\vspace{0.1cm}
  $\theta_{\circlearrowleft}$ & 0.589048619835 \tabularnewline\vspace{0.1cm}
  $\theta_{\asymp}^{(2)}$ & 0.785361375567 \tabularnewline
    \hline
    \vspace{0.15cm}
  $f_m$ & $O(10^{-16})$ \tabularnewline 
    $L$ &  $O(10^{-16})$ \tabularnewline 
  $T$ &  16.7 \tabularnewline
  $\alpha$ & 0.158 \tabularnewline 
\end{tabular}
  \label{thetaVals}
\end{table}


\section{Constraints on the ring exchange strength} \label{sec:Constraints}

We are constrained in our choices of parameters, as the relative sizes of the nearest-neighbour couplings, $J$ and ring-exchange coupling $J_{\circlearrowleft}$, are set by the ratio $\alpha = J_{\circlearrowleft}/J = \theta_{\circlearrowleft} /\theta_{\Box}$, since as we have seen in eq. (\ref{eqn:Iden1}) we end up turning on the  Hamiltonian $ {H}_{\times}+{H}_{\Box}+ \alpha {H}_{\circlearrowleft}$ for some time $t_{\Box} =t_{\circlearrowleft}=\theta_{\Box} \hbar /  J$. This means we are constrained to situations where we can set $\alpha$ to $\theta_{\circlearrowleft} /\theta_{\Box}$. We would expect $\alpha \lesssim 0.17$ (see e.g.~\cite{Katanin02,Coldea01,Mizel2004}), which is why we have chosen to use the particular parameter set shown in table~\ref{thetaVals} which has $\alpha = 0.158$.

However, we are not completely constrained to this value,  since all of the $\theta$ values are just phases. Thus we are free to add factors of $2 \pi$ to any of them, and we can add multiples of $2 \pi$ to $\theta_{\Box}$ in order to decrease the value of $\alpha$ in our implementation. Since $\theta_{\Box} = J t_{\Box}/\hbar$, the trade-off is that increasing $\theta_{\Box}$ corresponds to either increasing $J$ or increasing $t_{\Box}$, and under the assumption that we are using the strongest couplings possible, this really means increasing the gate time by $2\pi \hbar / J$ every time we add a multiple of $2 \pi$. This is not ideal, but at least gives us some more flexibility in our value of $\alpha$. 
Even using this method, we still seem to be tied down to a few precise values of $\alpha$. To get around this, we notice that we can split up the ring exchange into two parts:
\begin{align}
{U}_S &= \exp \left(-i {H}_{\times} \theta_{\Box} \right) \exp \left(-i {H}_{\Box} \theta_{\Box} \right) exp \left(-i {H}_{\circlearrowleft} \theta_{\circlearrowleft} \right) \nonumber \\
&= \exp \left(-i {H}_{\times} (\theta_{\Box}^a + \theta_{\Box}^b) \right) \exp \left(-i {H}_{\Box} (\theta_{\Box}^a + \theta_{\Box}^b)\right) \times ... \nonumber\\ &exp \left(-i {H}_{\circlearrowleft} (\theta_{\circlearrowleft}^a + \theta_{\circlearrowleft}^b) \right) \nonumber\\
&= \exp \left(-i {H}_{\times} \theta_{\Box}^a \right)\exp \left(-i {H}_{\Box} \theta_{\Box}^a \right) \exp \left(-i {H}_{\circlearrowleft} \theta_{\circlearrowleft}^a \right) \times... \nonumber\\
& \exp \left(-i {H}_{\times} \theta_{\Box}^b \right)\exp \left(-i {H}_{\Box} \theta_{\Box}^b \right) exp \left(-i {H}_{\circlearrowleft} \theta_{\circlearrowleft}^b \right) \nonumber \\
&= {U}_{\times}^a {U}_{\Box}^a {U}_{\circlearrowleft}^a  {U}_{\times}^b{U}_{\Box}^b {U}_{\circlearrowleft}^b \nonumber\\
&:={U}_S^a {U}_S^b 
\end{align}
so we now have two ring exchange interactions, ${U}_S^a$ and ${U}_S^b$ which have the same form as $H_S$ but with different values of $\alpha$. Now since $\theta_n := J_n t_n / \hbar$, and since ${U}_S^a{U}_S^b = {U}_S$ and all of the terms commute, this means that:
\begin{align}\label{eqn:Constr}
&J^a t_a + J^b t_b = \theta_{\Box} \nonumber\\
&J_{\circlearrowleft}^a t_a + J_{\circlearrowleft}^b t_b = \theta_{\circlearrowleft}.
\end{align}
where $\theta_{\Box}$, $\theta_{\circlearrowleft}$ are the parameters we found in the search in sec.~\ref{sec:Two}, and we have defined $\theta_{\Box}^a = J^a t_a$, $\theta_{\Box}^b = J^b t_b$. For simplicity, we take $J^a = J^b = J$, without loss of generality, since we are free to scale these parameters as we wish, provided we scale the corresponding $J_{\circlearrowleft}$ values correctly. Since $\theta_{\Box}$ and $\theta_{\circlearrowleft}$ are phases, we are free to add multiples of $2\pi$ to these values, so we replace $\theta_{\Box}$ with $\theta_{\Box}^{(n)}$, and rearrange equation (\ref{eqn:Constr}) to give
\begin{align}
&t_a = \frac{\theta_{\Box}^{(n)}}{J}\left[ \frac{\alpha^{(n)} - \alpha_b}{\alpha_a - \alpha_b} \right] \nonumber\\
&t_b = \frac{\theta_{\Box}^{(n)}}{J}\left[ \frac{\alpha_a- \alpha^{(n)} }{\alpha_a  - \alpha_b} \right],
\end{align}
where $\alpha_a := \theta_{\circlearrowleft}^a / \theta_{\Box}^a =J_{\circlearrowleft}^a / J^a $, $\alpha_b := \theta_{\circlearrowleft}^b / \theta_{\Box}^b = J_{\circlearrowleft}^b / J^b$, and $\alpha^{(n)}$ is defined above. For $t_a$ and $t_b$ to be positive, we need couplings such that $\alpha_a > \alpha^{(n)}  >\alpha_b$. So this tells us that if we are able to control the relative strengths of the ring and nearest-neighbour terms, and if we could get them such that $\alpha_a >\alpha^{(n)} >\alpha_b$ is satisfied, then regardless of what the actual values of $\alpha_a$ and $\alpha_b$ are, we can create the $CZ$ gate (at the expense of adding more interactions and thus increasing the time of the gate).


\section{Including noise}\label{sec:Performance}

We now look at the performance of this gate under the influence of noise. The two types of noise we consider are errors in coupling strengths or timing of the gates (i.e.\ random errors in the $\{ \theta \}$ parameters when implementing each of the Hamiltonians in (\ref{Usequence})) and fluctuations in magnetic fields acting on the qubits during the gate implementation (which would normally be protected by the decoherence-free or supercoherent properties, but are not while our gates are being implemented). The reasons for picking these particular types of errors are that one of the most promising systems in which to implement these encoded qubits is in arrays of quantum dots, and there have been several advances towards achieving these, e.g.~\cite{Laird2010,Thalineau2012}. These quantum dots are susceptible to errors in exchange coupling due to charge fluctuation~\cite{Burkard1999} (which we are modelling by fluctuations in the $\{ \theta \}$ values) and fluctuations in external magnetic field due to the nuclear spin bath or stray magnetic fields (see e.g.~\cite{Taylor2007}). When looking at these magnetic fluctuations, we look at two cases; in the first case we assume that magnetic field fluctuations are roughly uniform over each encoded qubit (so that there is collective decoherence acting on it, see Sec.~\ref{sec:Intro}), but the magnetic fields acting on different encoded qubits are different in magnitude and direction. In the second case, we consider magnetic field fluctuations acting independently on each physical qubit (i.e.\ a situation where a supercoherent qubit would be more appropriate). For all errors, we assume that the time scale for the fluctuations is large compared to the time to perform the gate, which is typically the case~\cite{Hu2006,Merkulov2002,Taylor2007}.

To measure the effects of noise, we calculate the gate fidelity using the techniques of quantum process tomography~\cite{Gilchrist2005,NielsenChuang}. Suppose we have a process acting on a state $\rho$ in a d-dimensional Hilbert space, such that $\rho$ is mapped to $\mathcal{E} ( \rho)$. We can describe this process as
\begin{equation}\label{eqn:Map}
\mathcal{E} (\rho) = \sum_i {E}_i \rho {E}_i^{\dagger}
\end{equation}
where $\{E_i \}$ are Kraus operators which satisfy $\sum_i {E}_i {E}_i^{\dagger} \le \idop$, with equality iff $\mathcal{E}$ is a trace-preserving map. In this case, the Kraus operators take the form
\begin{align}
E_1&= P e^{i\Phi} {V}^{\dagger}{U} {V} \nonumber\\
E_2&= (1-P) e^{i\Phi} {V}^{\dagger}{U} {V}
\end{align}
where $P$ is a projection into the logical subspace ($P = P_3 $ or $P_4$ depending on whether we are considering the 3- or 4-qubit encoding. For definitions of $P_3$ and $P_4$ see Sec.~\ref{sec:Two}), $V$ is a local rotation, $\Phi$ is a global phase, and $U$ is the unitary found in Sec.~\ref{sec:Two} which is locally equivalent to a $CZ$ gate. We expect the local operations to be easy to find, so we do not find these explicitly; instead we take the gate with zero noise ${U}_{tot}$ corresponding to the parameter set $\{ \theta \}$ in Sec.~\ref{sec:Two}, and find the matrix ${V}$ which diagonalises it (i.e. $e^{i\Phi} {V}^{\dagger} {U}_{tot} {V} = CZ$, where $\Phi$ accounts for any phase terms). Then once we have a gate with noise added, ${U}'$, we still use ${V}$ and $\Phi$ to convert ${U}'$ into a noisy $CZ$ gate (i.e. $e^{i\Phi} {V}^{\dagger}{U}' {V} \approx CZ$), since we are restricted to always using the same local operations (i.e.\ we cannot change our local operations since we don't know what the noise is doing to our system). 

Rather than finding the fidelity of the gate as a whole, it is more useful to decompose this noisy gate into two parts: the probability $1-L$ of being in the logical subspace (where $L$ is the leakage defined in Sec.~\ref{sec:Two}), and the fidelity of the gate after post-selecting on the measurement outcome, since we could in principle use a heralding scheme to reduce the effects of leakage out of the logical subspace. This also makes it easier to see what happens to the gate as noise is added. To measure the fidelity of the gate after post-selection we use the process fidelity~\cite{NielsenChuang}. To find this process fidelity, we decompose $E_1$ as ${E}_1= \sum_m a_{m} {A}_m$, where $\{ {A}_m \}$ is a set of $4 \times 4$ matrices that form an orthogonal basis under the Hilbert-Schmidt inner product, i.e. $tr({A}_m^{\dagger} {A}_n) = 4 \delta_{mn}$. Then the effect of the gate after post-selecting on the measurement is
\begin{align}
\mathcal{E} (\rho) &={E}_1 \rho {E}_1^{\dagger} \nonumber\\
&= \frac{1}{(1-L)}\sum_{mn} (a_{m}a_n^{*}) {A}_m \rho {A}_n^{\dagger} \nonumber\\
&\equiv \frac{1}{(1-L)}\sum_{mn} \chi_{mn} {A}_m \rho {A}_n^{\dagger},
\end{align}
where $\chi_{mn} := a_{m}a_n^{*}$ is the `process matrix', and we divide by $(1-L)$ to account for the probabilistic nature of the measurement. Then we can absorb this into $\chi$ by letting $\chi/(1-L)\to \chi$. We can then compare this normalised process matrix $\chi$ to the process matrix of an ideal $CZ$ gate, $\chi_{id}$, using the process fidelity $F_p$~\cite{NielsenChuang}:
\begin{align}
F_p(\chi,\chi_{CZ}) =\textrm{tr}\left( \sqrt{ \sqrt{\chi_{id}} \chi \sqrt{\chi_{id}}} \right)^2.
\end{align}
$(1-(1-L)F_{p})$ has the interpretation as the upper bound of the average failure probability $\bar{p}_e$~\cite{Gilchrist2005}, making this a natural choice for measuring the accuracy of the gate. Note that a subtlety arises when dealing with the 3-qubit DF subsystem, in that $P_3$ projects onto a 16-dimensional space, so we cannot directly compare this to the 4-dimensional $CZ$ gate. To get around this, we can instead project onto one of the gauge degrees of freedom, so for instance we could replace $P_3$ with $P_3^{(1,1)}$ (defined in Sec.~\ref{sec:Two}),
provided $P_3^{(1,1)} U P_3^{(1,1)} \ne 0$, otherwise we could project onto any of the other 3 gauge subspaces. Then instead of normalising $\chi$ by dividing by $(1-L)$, we can just divide by $\textrm{tr}(\chi)$ since $\textrm{tr}(E_1 \rho E_1) = \textrm{tr}(\chi)$, and the effect is the same. Note that the leakage is still calculated in the same way as in Sec.~\ref{sec:Two}, independently of the method used to project into the 4-dimensional subspace.

\subsection{Coupling Errors}

\begin{figure}[h]
\centering\includegraphics[width = 20pc]{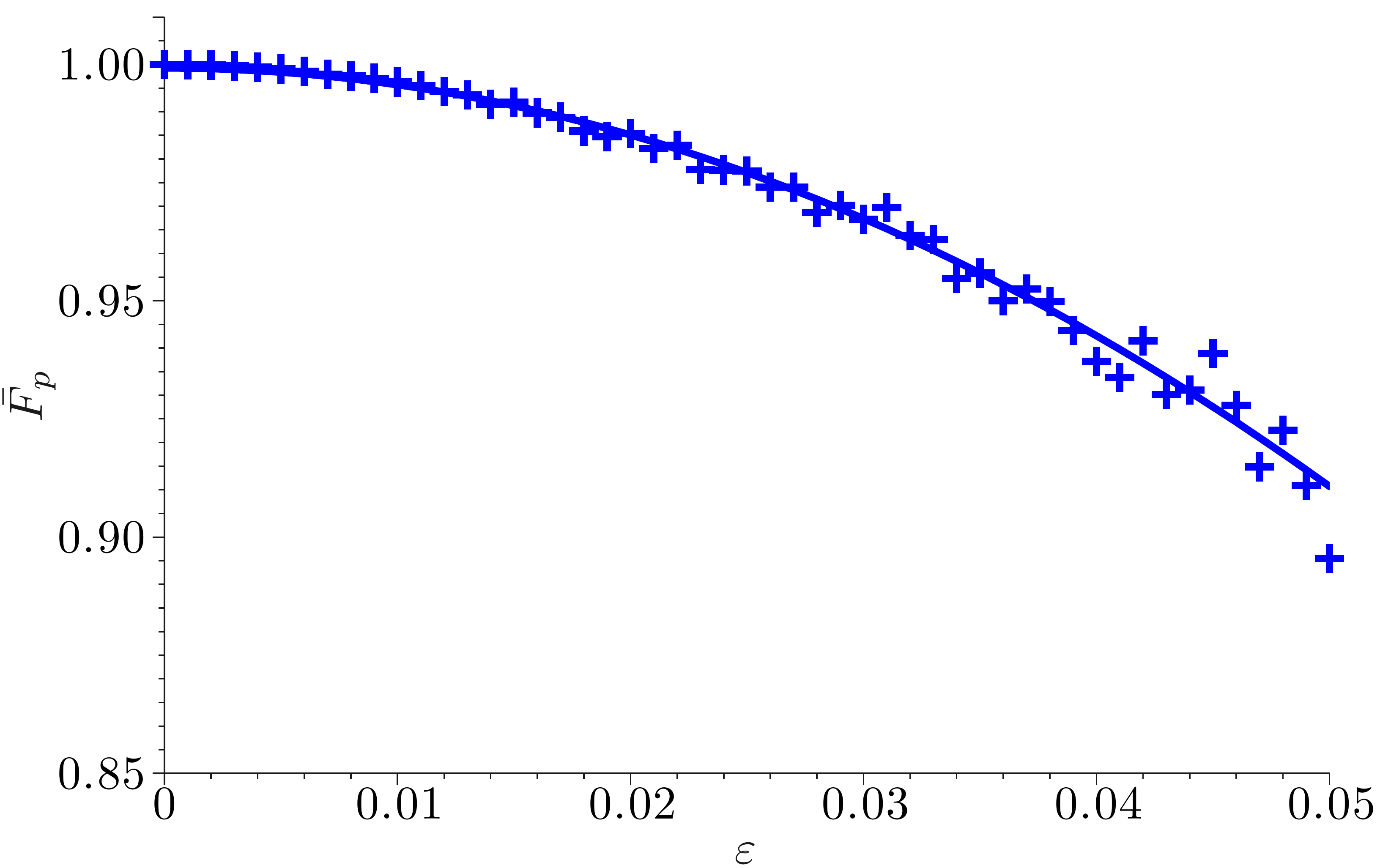}
\caption{The average process fidelity $\bar{F}_p$ when performing the gate with random fluctuations in gate times, where the fluctuations have standard deviation $\varepsilon$. Only results for the 4-qubit encoding are shown, since both encodings showed a similar behaviour. A fit to a curve of the form $y = 1 - c \varepsilon^2$ is also shown, with $c=35.4$.}
\label{TimingErrors}
\end{figure}

To simulate random errors in values of the couplings between physical qubits or timing when implementing gates, we added random Gaussian noise to each of the $\theta_n$ parameters:
\begin{equation}
\theta_n \rightarrow \theta_n + \delta\theta_n,
\end{equation}
where $\delta\theta_n$ is sampled from a Gaussian distribution with mean 0 and standard deviation $\varepsilon$.  An example of one cause of such errors is charge fluctuations in quantum dot systems (see e.g.~\cite{Burkard1999}).
Over the range $\varepsilon \in [0,0.05]$ of $\varepsilon$, we calculated the process fidelity over 250 iterations taken from a normal distribution with standard deviation $\varepsilon$ and mean 0, finding the average over all of these iterations (note that all the interactions commute with ${S}^2$, so we can confine ourselves to a subspace of constant $S$). The results are shown in Fig.~\ref{TimingErrors}, with only one set of results shown since both the 3- and 4-qubit encoding give very similar results. The process fidelity falls off slowly and stays above $0.9$ for $\epsilon \lesssim 0.05$. A reasonable estimate of these fluctuations in gate couplings would be around $0.01$~\cite{Burkard1999}, at which point both gates have fidelity $\sim 0.99$, so we can see that these gates still have high fidelity even with this level of noise. Over the entire range of $\epsilon$, the leakage stayed below $0.003$, and the leakage at $\epsilon \sim 0.01$ is around $O(10^{-6})$. So overall we can achieve an overall average gate failure probability of $\bar{p}_e \lesssim 0.01$ even with a reasonable level of coupling errors.

\subsection{Magnetic fluctuations}

 \begin{figure}[b]
  \centering\includegraphics[width = 20pc]{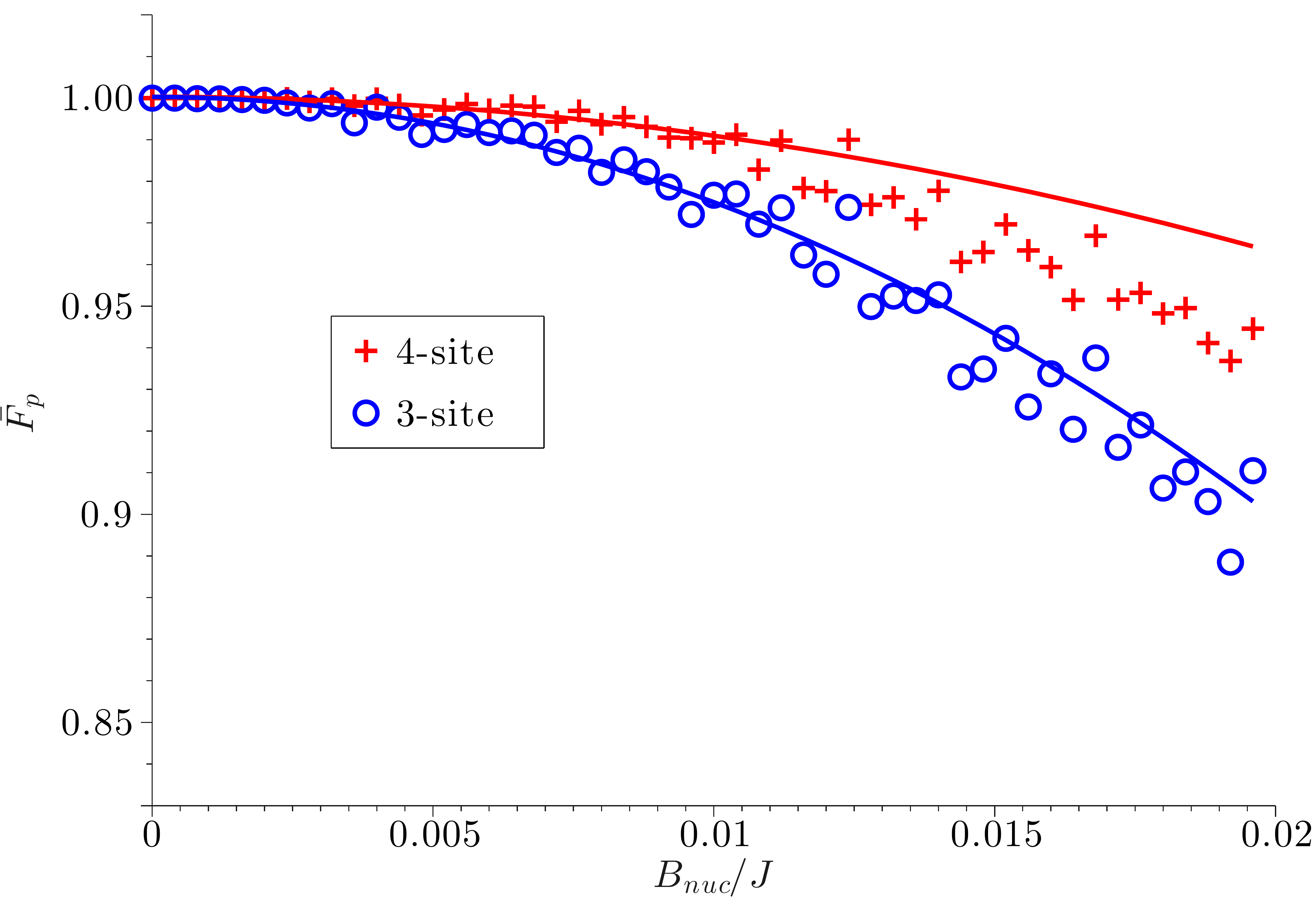}
 \caption{Average process fidelity $\bar{F}_p$ as a function of $B_{nuc}/J$, the magnetic field strength, with the fields pointing in random directions on each logical qubit. The range is over $B_{nuc}/J = 0 \rightarrow 0.02$, and results corresponding to a 3-qubit or 4-qubit encoding are both shown. A fit of all data up to $B_{nuc}/J = 0.01$ to a curve of the form $y = 1 - c \varepsilon^2$ is also shown, with $c_4=95.6$, $c_3=252.9$ for the 4-qubit and 3-qubit encoding respectively. Note that we use this fit to see roughly when the fidelity starts to deviate from a quadratic decay, not as a best fit to the data.}
 \label{MagEvolveLong}
 \end{figure}
 
 \begin{figure}[b]
  \centering\includegraphics[width = 20pc]{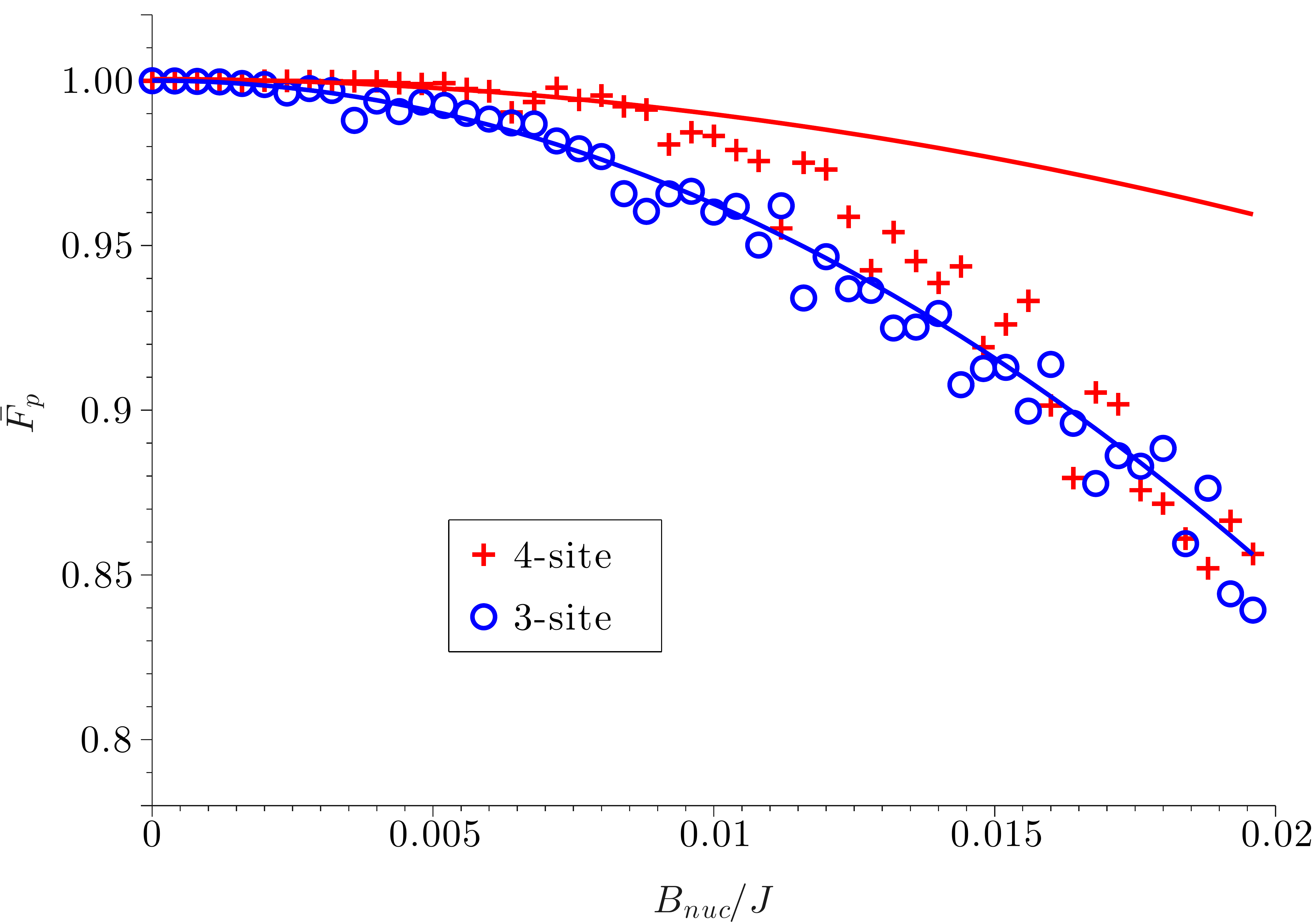}
 \caption{Average process fidelity $\bar{F}_p$ as a function of $B_{nuc}/J$, the magnetic field strength relative to the exchange coupling, with the fields pointing in random directions on each physical qubit. The range is over $B_{nuc}/J = 0 \rightarrow 0.02$, and results corresponding to a 3-qubit or 4-qubit encoding are both shown. A fit of all data up to $B_{nuc}/J = 0.01$ to a curve of the form $y = 1 - c \varepsilon^2$ is also shown, with $c_4=106.9$, $c_3=374.7$ for the 4-qubit and 3-qubit encoding respectively. Note that we use this fit to see roughly when the fidelity starts to deviate from a quadratic decay, not as a best fit to the data.}
 \label{MagEvolveSingle}
 \end{figure}

If we were to implement this supercoherent qubit in a quantum dot, then there could be random magnetic field fluctuations due to the nuclear spins in the substrate material, or stray magnetic fields. We studied two scenarios which could occur; in one case we looked at the effects of having random field fluctuations which are uniform over all physical qubits, but which may vary between encoded qubits (so that the collective decoherence assumption is valid for single encoded qubits, but when we interact two of these the assumption is not valid). In the other case we considered having independent random magnetic field fluctuations on each physical qubit. For both of these cases, we followed the arguments in~\cite{Merkulov2002, Taylor2007}, using the quasi-static approximation in which the magnetic field from the nuclei $\vec{B}$ stays constant over the time we perform the gate, has random direction and has magnitude $| \vec{B} |$ following a Gaussian probability distribution
\begin{align}
P(| \vec{B} |) = \frac{1}{(2 \pi  B_{nuc}^2)^{3/2}} \exp(- | \vec{B} |^2 / 2B_{nuc}^2),
\end{align}
 where $B_{nuc}$ is the standard deviation in fluctuations of magnetic field. We took the average over 250 iterations, with each iteration having a different magnitude and direction of magnetic field sampled from the above Gaussian distribution. The magnetic field strength was taken relative to the nearest-neighbour coupling strength $J$. In each case we found how the process fidelity varied for the 3-qubit and 4-qubit encoding. The results are shown in Fig.~\ref{MagEvolveLong} and Fig.~\ref{MagEvolveSingle}, together with a fit of all data up to $B_{nuc}/J = 0.01$ to a curve of the form $\bar{F}_p = 1 - c \varepsilon^2$ (we use this fit to see roughly when the fidelity starts to deviate from a quadratic decay, not as a best fit to the data). 
 
 Based on the exchange values given in~\cite{Loss1998} and the values for $B_{nuc}$ given in~\cite{Taylor2007}, we would expect $B_{nuc}/J$ to be at most $\sim 0.01$. With this kind of nuclear field present, the 4-qubit encoding achieves a process fidelity of $\bar{F}_p \sim 0.99$ and a leakage probability of $L \sim 0.002$ if there are errors across the encoded qubits, or $(\bar{F}_p \sim 0.97, L \sim 0.002)$ if there are errors on each individual physical qubit. The 3-qubit encoding achieves $(\bar{F}_p\sim 0.97,L \sim 0.002)$ if there are errors across encoded qubits, or $(\bar{F}_p \sim 0.95,L \sim 0.002)$ if there are errors on physical qubits. 
 
So overall, as we might expect, the qubits are much more robust to errors which are uniform on the physical qubits, but less robust to errors which vary between physical qubits. Also, the gate on the 3-qubit encoding performs slightly worse than the 4-qubit one, which we are currently unable to explain. Using a 4-qubit encoding in the presence of magnetic fluctuation across encoded qubits, with a strength we might expect in a realistic system, we can achieve process fidelities of $\bar{F}_p > 0.99$ and leakages of $L \sim 0.002$ (leading to an overall average gate failure probability of $\bar{p}_e \lesssim 0.01$). The results for the situation with errors on physical qubits are unsurprisingly worse, but we can still achieve $\bar{F}_p \sim 0.97$, $L \sim 0.002$, giving an overall average gate failure probability of $\bar{p}_e \lesssim 0.03$. These results could be improved if we reduced the effects of fluctuations in nuclear spin, such as the methods presented in~\cite{Imamoglu2003} and~\cite{Reilly2008}.


\section{Conclusions}
We have demonstrated a simple way to implement a controlled-$Z$ gate in the 4-qubit decoherence-free subspace and the 3-qubit decoherence-free subsystem, using a sequence of 5 operations (excluding local operations), and including ring exchange interactions. The gate we have found minimises the Makhlin invariant function $f_m$ to within machine precision, suggesting the existence of an exact solution. We introduced errors when performing these gates, to simulate errors in coupling strength or gate times, and to simulate fluctuations in magnetic field due to some external environment, e.g.\ nuclear spins in a quantum dot. We found that the 4-qubit gate maintained an average failure probability of $\bar{p}_e \lesssim 0.01$ even with nuclear fluctuations of around $1\%$ of $J$ over the encoded qubits, or with timing errors of up to around $1 \%$ of $J/\hbar$, where $J$ is the strength of the nearest-neighbour exchange coupling.

Such a gate could be useful in systems where ring exchange is particularly prominent, or in situations where it is particularly important to keep the number of pulses to a minimum, or where the control is limited. It also demonstrates that perhaps ring exchange can be used as a resource to produce simplified gates, which we might intuitively expect since a direct $CZ$ gate on these encoded qubits would involve a four-body interaction, which is present in the ring exchange terms. In future we would like to investigate this more rigorously, to see if gates on these encoded qubits involving ring exchange terms always outperform gates without ring exchange interactions. We would also like to perform searches using the more general form of the ring exchange interaction rather than the one we have used here, and also including the effects of magnetic flux on the couplings as reported in~\cite{Scarola2005}.  It would also be interesting to see if we could extend the techniques such as dynamical decoupling and leakage reduction, previously applied to 3-qubit encoded qubits in~\cite{West2012,Fong2011}, to this gate, to improve the performance in the presence of noise.

\begin{acknowledgments}
SB is supported by an ERC starting grant. BA is supported by the EPSRC. We thank D.P.DiVincenzo for pointing out refs.~\cite{Fong2011,West2012}.
\end{acknowledgments}

\appendix

\section{Equivalence of gates for 3 and 4-qubit encoding}\label{App1}
Here we briefly discuss which pulse sequences that realise a $CZ$ gate (excluding local rotations) in the 3-qubit DF subsystem also realise a $CZ$ gate in the 4-qubit DF subspace, and vice-versa. To begin with, we restate the definitions made in Sec.~\ref{sec:Background} of the states involved in the three qubit DF subsystem
\begin{align}
&\ket{\bar{0}^{(3)}_{+1}}=\ket{\psi^-}_{12}\ket{0}_3\nonumber\\
&\ket{\bar{0}^{(3)}_{-1}}=\ket{\psi^-}_{12}\ket{1}_3\nonumber\\
&\ket{\bar{1}^{(3)}_{+1}}=\frac{1}{\sqrt{3}}( \sqrt{2} \ket{T_+}_{12}\ket{1}_3 - \ket{T_0}_{12}\ket{0}_3)\nonumber\\
&\ket{\bar{1}^{(3)}_{-1}}=\frac{1}{\sqrt{3}}( \ket{T_0}_{12}\ket{1}_3 - \sqrt{2} \ket{T_{-}}_{12}\ket{0}_3),
\end{align}
and for the 4-qubit DF subspace:
\begin{align}\label{eqn:4DFS}
\ket{\bar{0}^{(4)}} & :=  \ket{\psi^-}_{12} \ket{\psi^-}_{34}\nonumber\\
\ket{\bar{1}^{(4)}} & :=  \frac{1}{\sqrt{3}} \left[ \ket{T_+}_{12} \ket{T_-}_{34} - \ket{T_0}_{12} \ket{T_0}_{34}\right.\nonumber\\ 
&\left.+ \ket{T_-}_{12} \ket{T_+}_{34}\right].
\end{align}
 Observe that we can rewrite the states in (\ref{eqn:4DFS}) in this form:
\begin{align}\label{eqn:App1}
\ket{\bar{0}^{(4)}}& = \frac{1}{\sqrt{2}} \left[ \ket{\bar{0}^{(3)}_{+1}}\ket{1} - \ket{\bar{0}^{(3)}_{-1}}\ket{0} \right]\nonumber\\
\ket{\bar{1}^{(4)}} & = \frac{1}{\sqrt{2}} \left[ \ket{\bar{1}^{(3)}_{+1}}\ket{1} - \ket{\bar{1}^{(3)}_{-1}} \ket{0} \right].
\end{align}

A valid pulse sequence for the 3-qubit DF subsystem will perform a gate which is (locally equivalent to) a $CZ$ gate on two 3-qubit states up to a gauge transformation. In general, this gauge transformation may mean that this pulse sequence does not work for the 4-qubit DF subspace (it may not amount to a simple local rotation in the 4-qubit case). However, in certain situations, a valid pulse sequence for the 3-qubit DF subsystem will work. For example, since exchange interactions commute with $S$ and $S_z$ over all the qubits, any interactions made up of exchange coupling only (such as the gate found in~\cite{Fong2011}) will not couple different gauge states. So certain pulse sequences for the 3-qubit DF subsystem (excluding local rotations) enables us to perform a $CZ$ gate on the 4-qubit DF subspace. 

The converse is also not necessarily true; a valid pulse sequence on the 4-qubit DF subspace will not necessarily work on the 3-qubit DF subsystem, simply because in the 4-qubit case, interactions can be over 8 physical qubits rather than 6. However, if all interactions are confined to 3 physical qubits on each logical qubit (which is the case in our protocol) then a valid pulse sequence for the 4-qubit DF subspace is also a valid gate on the 3-qubit DF subsystem. This can easily be seen since any gate locally equivalent to a $CZ$ gate is also locally equivalent to the interaction $\bar{Z}_A \bar{Z}_B$ acting between two logical qubits $A$ and $B$, where $\bar{Z}_k$ is the logical $Z$ operator acting on logical qubit $k$. Since the logical $Z$ operator for the 4-qubit encoding is the same as for the 3-qubit encoding (an exchange interaction between two qubits, e.g. $E_{1,2}$ using the above definitions of the states), then this gate is also locally equivalent to a $CZ$ gate for the 3-qubit encoding.

\bibliography{DFSpaper}

\end{document}